\documentclass[%
reprint,
runinaddress,
superscriptaddress,
amsmath,amssymb,
aps,
prb,
]{revtex4-2}

\usepackage{graphicx}
\usepackage{dcolumn}
\usepackage{bm}
\usepackage{float}

\graphicspath{ {./images} }

\begin{document}
	
	\preprint{APS/123-QED}
	
	\title{Bias-field control of the Néel skyrmion nonlinearity in a confined nanostructure}
		\author{A.V. Valkov}
	\email{valkov.valkovalex@yandex.ru}
	\affiliation{Kotel’nikov Institute of Radio-Engineering and Electronics of RAS, Moscow 125009, Russia} 
	\affiliation{Moscow Institute of Physics and Technology (National Research University), Dolgoprudnyi, Moscow region, 141701 Russia}
	
	\author{A.A. Matveev}
	\email{maa.box@yandex.ru}
	\affiliation{Kotel’nikov Institute of Radio-Engineering and Electronics of RAS, Moscow 125009, Russia} 
	\affiliation{Moscow Institute of Physics and Technology (National Research University), Dolgoprudnyi, Moscow region, 141701 Russia}
	\author{O.Yu. Arkhipova}
	\affiliation{Kotel’nikov Institute of Radio-Engineering and Electronics of RAS, Moscow 125009, Russia} 
	\affiliation{Moscow Institute of Physics and Technology (National Research University), Dolgoprudnyi, Moscow region, 141701 Russia}
	\affiliation{Bauman Moscow State Technical University, Moscow, 105005, Russia}
	\author{R.V. Shcherbakov}%
	\affiliation{Kotel’nikov Institute of Radio-Engineering and Electronics of RAS, Moscow 125009, Russia} 
	\affiliation{National Research University Moscow Power Engineering Institute, Moscow, 111250 Russia}%
	\author{A.R. Safin}%
	\affiliation{Kotel’nikov Institute of Radio-Engineering and Electronics of RAS, Moscow 125009, Russia} 
    \affiliation{Moscow Institute of Physics and Technology (National Research University), Dolgoprudnyi, Moscow region, 141701 Russia}
	\affiliation{National Research University Moscow Power Engineering Institute, Moscow, 111250 Russia}%
	\author{S.A. Nikitov}%
\affiliation{Kotel’nikov Institute of Radio-Engineering and Electronics of RAS, Moscow 125009, Russia} 
\affiliation{Moscow Institute of Physics and Technology (National Research University), Dolgoprudnyi, Moscow region, 141701 Russia}
\affiliation{Laboratory of Magnetic Metamaterials, Saratov State University, 410012 Saratov, Russia}
	
	\date{\today}
	
\begin{abstract}
We study the nonlinear dynamics of a cylindrical skyrmion-based oscillator within the
framework of a generalized Thiele model. We demonstrate the influence of an bias magnetic field applied perpendicular to the plane of the
nanocylinder on the oscillation frequency and on the nonlinearity coefficient.
It is shown that the field tunes the response frequency, while also producing a substantial change in the nonlinear frequency shift. We find that the nonlinearity coefficient
reverses its sign as the field crosses a certain critical value. The variation
of the nonlinearity coefficient with the field is clearly illustrated by the
observed qualitative changes in nonlinear amplitude--frequency responses. Controlling the nonlinear properties of a skyrmion oscillator by changing the bias magnetic field opens up prospects for creating tunable  computational elements for neuromorphic applications.
\end{abstract}

\maketitle

\section{INTRODUCTION}
Research on ferromagnetic materials in spintronics is aimed at exploring their potential for the development of next-generation microwave devices for the reception, transmission, and processing of information~\cite{nikitov2020dielectric,shao2021roadmap,matasov2025effect,locatelli2014spintorque,ustinov2026digit}. Such materials are considered to be of use for both classical delay lines, phase shifters, and logic elements~\cite{serga2010yig,khitun2010magnonic} and for elements of quantum transducers~\cite{lachancequirion2019hybrid,samoilenko2025magnon,tabuchi2015coherent}. Moreover, a lot of applications rely on the nonlinear properties of magnetization dynamics~\cite{ref53,louis2017low}. However, the most promising magnetic materials are those in which inhomogeneous magnetization states, such as skyrmions, can be stabilized.

A magnetic skyrmion is a non-trivial vortex-like magnetization distribution, which is commonly characterized by a topological charge taking the values $\pm 1$~\cite{ref1,ref2,ref3}. Such a structure may be regarded as consisting of three regions: the core, in which the magnetization vector $\bm{M}$ is perpendicular to the film plane; a transition region; and an outer region, in which the magnetization is antiparallel to the direction of $\bm{M}$ in the core. Magnetization within this transition region rotates continuously. This rotation may occur either along the radial direction, in which case the skyrmion is said to be of N\'eel type, or along the direction perpendicular to the radial one, which corresponds to Bloch-type skyrmions. Such structures can be stabilized in the presence of the Dzyaloshinskii--Moriya interaction (DMI)~\cite{ref2,ref3,ref4,ref5}, which in bulk non-centrosymmetric crystals with the B20 structure is referred to as bulk DMI and stabilizes the Bloch skyrmion. N\'eel skyrmions can arise in thin films, where they are stabilized by the interfacial DMI. 
Skyrmions have been observed in a variety of materials, including the bulk non-centrosymmetric crystals $\mathrm{MnSi}$, $\mathrm{FeCoSi}$ and $\mathrm{FeGe}$~\cite{ref6,ref7,ref8,ref9}, multilayer structures based on $\mathrm{Pt/Co/MgO}$, $\mathrm{Ta/CoFeB/MgO}$, $\mathrm{Ir/Co/Pt}$ and others~\cite{ref10,ref11,ref12,ref13,ref14,ref15,ref16,ref17,ref18}, as well as in multiferroics~\cite{ref19,ref20} and frustrated magnets~\cite{ref3,ref21}. Besides this broad materials base, a large number of techniques for generating skyrmion states are available, including laser-pulse heating~\cite{ref13,ref22}, mechanical strain~\cite{ref23}, injection of a spin-polarized current through the sample~\cite{ref24,ref25}, the localized magnetic field of a magnetic force microscope probe~\cite{ref12}, and electric-field pulses~\cite{ref26}; the methods can also be used to switch the skyrmion state, which is attractive for spintronic devices. 

The practical applications of skyrmions are varied. For example, these magnetization configurations are considered as information carriers in racetrack memory elements~\cite{ref1,ref2,ref3,ref27,ref28,ref29,ref30}, with logic operations implemented on them~\cite{ref31,ref32,ref33,ref34}, including basic gates and adders; as transistors~\cite{ref35}, in which the transmission of a skyrmion through the gate is governed by the applied voltage; and as the principal building blocks of neuromorphic computing units~\cite{ref36} possessing synaptic plasticity for both short-term and long-term memory functions. In addition, skyrmions have been observed in the free layer of a magnetic tunnel junctions ~\cite{ref10,ref11,ref37,wintz}. The presence of such a skyrmion state opens the way to designing multilevel magnetoresistive memory cells, caused by additional resistance that it provides, as well as spin-transfer nano-oscillators (STNOs) ~\cite{ref25,ref38,ref39}, which can operate efficiently when the so-called gyrotropic and breathing modes are excited~\cite{ref40,ref41}. It has been shown that such STNOs require a lower out-of-plane spin-polarized current density to sustain self-oscillations than is required in STNOs with a uniform magnetization state~\cite{ref39}.

Many studies have been devoted to the impact of the constant external magnetic field $B_0$ applied to such STNOs. The tuning of the frequencies of the gyrotropic and breathing skyrmion modes by varying $B_0$ has been studied in detail~\cite{ref42}. However, from the standpoint of practical applications, tuning the nonlinearity of spintronic oscillators is crucial. For example, tuning the nonlinearity of magnetic tunnel junctions has been used to generate neuromorphic hardware training~\cite{ref43}, and in STNOs the degree of chaos and the spiking waveform can be continuously tuned via the bias current~\cite{ref44}. This paper investigates the gyrotropic-mode frequency and the nonlinearity coefficient of a skyrmion oscillator dependencies on the magnitude of the bias magnetic field. To this end, in Section~\ref{sec:model} we describe the magnetization in material as a weighted sum of the skyrmion ansatz and a term describing the magnetization distribution at the sample boundary. In addition, we consider the deformation of the skyrmion shape that arises from its interaction with the boundary upon displacement from the equilibrium position. In Section~\ref{sec:dynamics} we use this model to derive a generalized Thiele equation describing the skyrmion dynamics with these deformations, and in Section~\ref{sec:hamiltonian} we recast it in Hamiltonian form to obtain the nonlinearity coefficient. In Section~\ref{sec:results} presents the resulting frequency and nonlinearity dependence on the bias magnetic field, compared against nicromagnetic simulations, and Section~\ref{sec:conclusion} summarizes our conclusions.
\section{PHYSICAL STRUCTURE AND
		MATHEMATICAL MODEL}
\label{sec:model}

\begin{figure}[t]
  \centering
  \includegraphics[scale = 1.4]{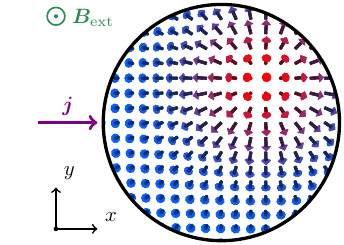}
  \caption{Sketch of the structure under study. The ground magnetization state is a N\'eel skyrmion, whose dynamics is excited by passing a spin-polarized current through the sample.}
  \label{fig:schematic}
\end{figure}

The structure under study is a thin nanocylinder with radius $R_0 = 50$~nm and thickness $l_z = 0.8$~nm, which in practice can be made of, for example, $\mathrm{Pt/Co}$ or $\mathrm{Pt/Co/Ir}$ (see Fig.~\ref{fig:schematic}). We consider a N\'eel skyrmion as the ground magnetization state of the sample. This state is stable owing to the presence of the interfacial DMI.

Let us construct a mathematical model of the magnetization distribution $\bm{M}(\bm{r})$ in the structure under study, considering the influence of the DMI which, as established in Ref.~\cite{ref29}, gives rise to magnetization inhomogeneities at the sample boundary. Following the approach previously used in Ref.~\cite{ref29}, we presented the magnetization of the sample as the weighted sum of the skyrmion ansatz $\bm{m}_{\mathrm{sk}}$ and a partially expelled N\'eel-type domain wall $\bm{m}_{\mathrm{dw}}$. This wall describes the magnetization distribution at the sample boundary and has to satisfy the Dzyaloshinskii--Moriya boundary condition~\cite{ref4,ref5}
\begin{equation} \frac{\partial \bm{M}}{\partial \bm{n}} = \frac{D_{\mathrm{DMI}}}{2A_{\mathrm{ex}}} \bigl[\,[\mathbf{e}_z,\bm{n}],\bm{M}\,\bigr], \label{eq:bc} \end{equation}
where $\bm{n}$ is the outward normal vector, $D_{\mathrm{DMI}}$ is the interfacial Dzyaloshinskii--Moriya constant and $A_{\mathrm{ex}}$ is the exchange stiffness constant.

We present $\bm{m}_{\mathrm{sk}}$ and $\bm{m}_{\mathrm{dw}}$ using spherical coordinates as
\begin{equation}
  \label{eq:msph}
  \begin{split}
    \bm{m}_{\mathrm{sk,dw}} &=
    \bigl(
      m^x_{\mathrm{sk,dw}},\; m^y_{\mathrm{sk,dw}},\; m^z_{\mathrm{sk,dw}}
    \bigr), \\
    m^x_{\mathrm{sk,dw}} &= \cos\varphi_{\mathrm{sk,dw}}\sin\theta_{\mathrm{sk,dw}}, \\
    m^y_{\mathrm{sk,dw}} &= \sin\varphi_{\mathrm{sk,dw}}\sin\theta_{\mathrm{sk,dw}}, \\
    m^z_{\mathrm{sk,dw}} &= \cos\theta_{\mathrm{sk,dw}}.
  \end{split}
\end{equation}
where $\varphi_{\mathrm{sk,dw}}$ and $\theta_{\mathrm{sk,dw}}$ are the azimuthal and polar angles of the local magnetization vector, respectively. For the skyrmion these angles can be found following relations
\begin{subequations}
  \label{eq:angles}
  \begin{gather}
    \theta_{\mathrm{sk}} = \pi - 2\arctan\!
      \left(\frac{\sinh(\varepsilon/\delta)}
                 {\sinh\bigl(r_{\mathrm{sk}}/\lambda\delta\bigr)}\right),
      \label{eq:angles-a} \\[4pt]
    \cos\varphi_{\mathrm{sk}} = \frac{X}{r_{\mathrm{sk}}},
    \sin\varphi_{\mathrm{sk}}  = \frac{Y}{r_{\mathrm{sk}}}.
      \label{eq:angles-b}
  \end{gather}
For the domain wall we use the ensuing expressions, adapted from Ref.~\cite{ref29}
  \begin{gather}
    \theta_{\mathrm{dw}} = \arccos\!
      \left(\tanh\frac{|\bm{r}| - r_c}{\lambda}\right),
      \label{eq:angles-c} \\[4pt]
    \cos\varphi_{\mathrm{dw}} =
      \operatorname{sign}(D_{\mathrm{DMI}})\frac{x}{|\bm{r}|},
    \sin\varphi_{\mathrm{dw}} =
      \operatorname{sign}(D_{\mathrm{DMI}})\frac{y}{|\bm{r}|}.
      \label{eq:angles-d}
  \end{gather}
\end{subequations}

In Eqs.~\eqref{eq:angles-a}--\eqref{eq:angles-d} the following notation is used: $\bm{r} = (x,y)$ is the radius vector of the point in space at which the sine and cosine of the angles $\varphi$ and $\theta$ are evaluated; $\bm{R} = (X,Y)$ are the Cartesian coordinates of the skyrmion guiding center; $\lambda = \sqrt{A_{\mathrm{ex}}/K_0}$ is the characteristic length scale; $r_c = R_0 + \lambda\operatorname{arcsech}D_0$ is the position of the center of the N\'eel-type domain wall, which lies outside the sample; $D_0 = |D_{\mathrm{DMI}}|/\sqrt{A_{\mathrm{ex}} K_0}$; $K_0 = K_u - \mu_0 M_s^2/2$, where $K_u$ is the perpendicular anisotropy constant, $\mu_0$ is the magnetic permeability and $M_s$ is the saturation magnetization; $r_{\mathrm{sk}} = \sqrt{(x-X)^2 + (y-Y)^2}$ is the distance from the skyrmion center to the point in question; and $\varepsilon$ and $\delta$ are the minimization parameters.

To describe the interaction of the skyrmion with the inhomogeneity at the sample boundary, we write the magnetization vector as the weighted sum
\begin{subequations}
  \label{eq:weighted}
  \begin{gather}
    \bm{M} = M_s\,
      \frac{\vartheta_{\mathrm{sk}}\bm{m}_{\mathrm{sk}}
            + \vartheta_{\mathrm{dw}}\bm{m}_{\mathrm{dw}}}
           {\bigl|\vartheta_{\mathrm{sk}}\bm{m}_{\mathrm{sk}}
            + \vartheta_{\mathrm{dw}}\bm{m}_{\mathrm{dw}}\bigr|},
      \label{eq:weighted-a} \\[4pt]
    \vartheta_{\mathrm{sk}} =
      \frac{|\theta_S - \theta_{\mathrm{sk}}|}
           {|\theta_S - \theta_{\mathrm{sk}}| + (1-|\bm{R}|/R_0)|\theta_S - \theta_{\mathrm{dw}}|},
      \label{eq:weighted-b} \\[4pt]
    \vartheta_{\mathrm{dw}} =
      \frac{(1-|\bm{R}|/R_0)|\theta_S - \theta_{\mathrm{dw}}|}
           {|\theta_S - \theta_{\mathrm{sk}}| + (1-|\bm{R}|/R_0)|\theta_S - \theta_{\mathrm{dw}}|}.
      \label{eq:weighted-c}
  \end{gather}
\end{subequations}
where $\vartheta_{\mathrm{sk,dw}}$ are the weighting factors, whose values are proportional to the difference between the polar angle in the uniformly magnetization state of the sample, $\theta_S = \pi$, and $\theta_{\mathrm{sk,dw}}$.

\begin{figure}[t]
  \centering
  (a)\\[0pt]
  \includegraphics[width=\linewidth]{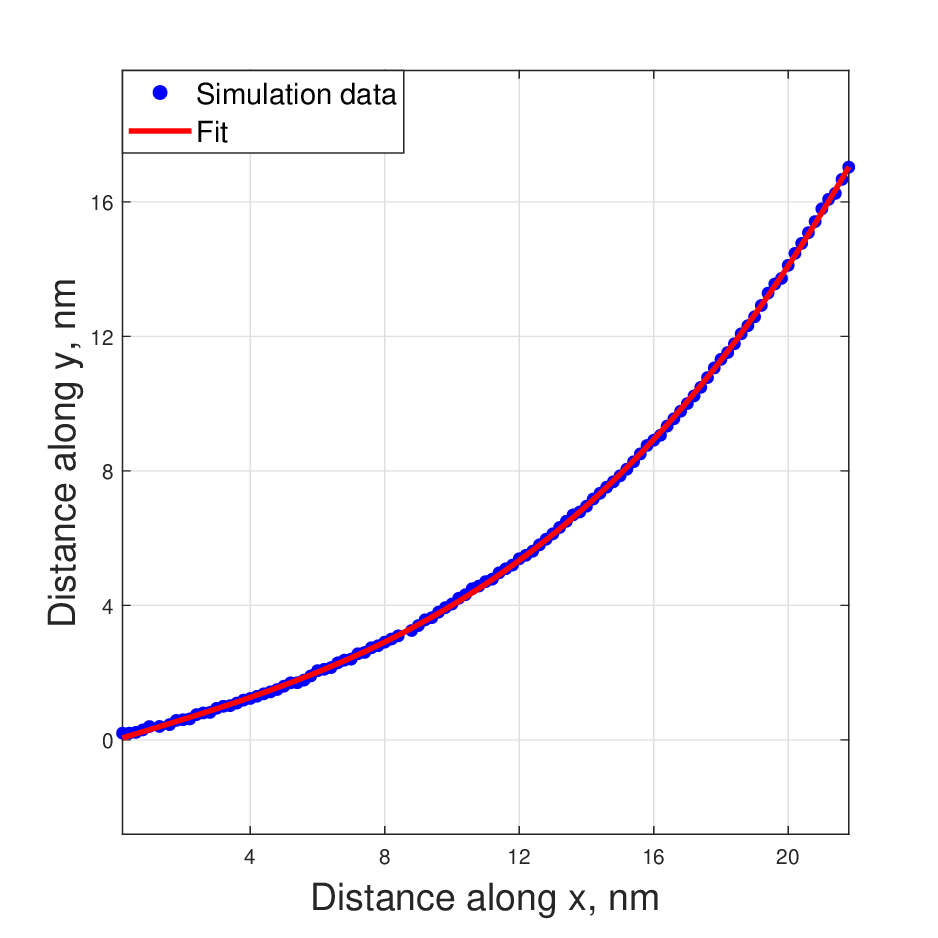}\\[0pt]
  (b)\\[0pt]
  \includegraphics[width=\linewidth]{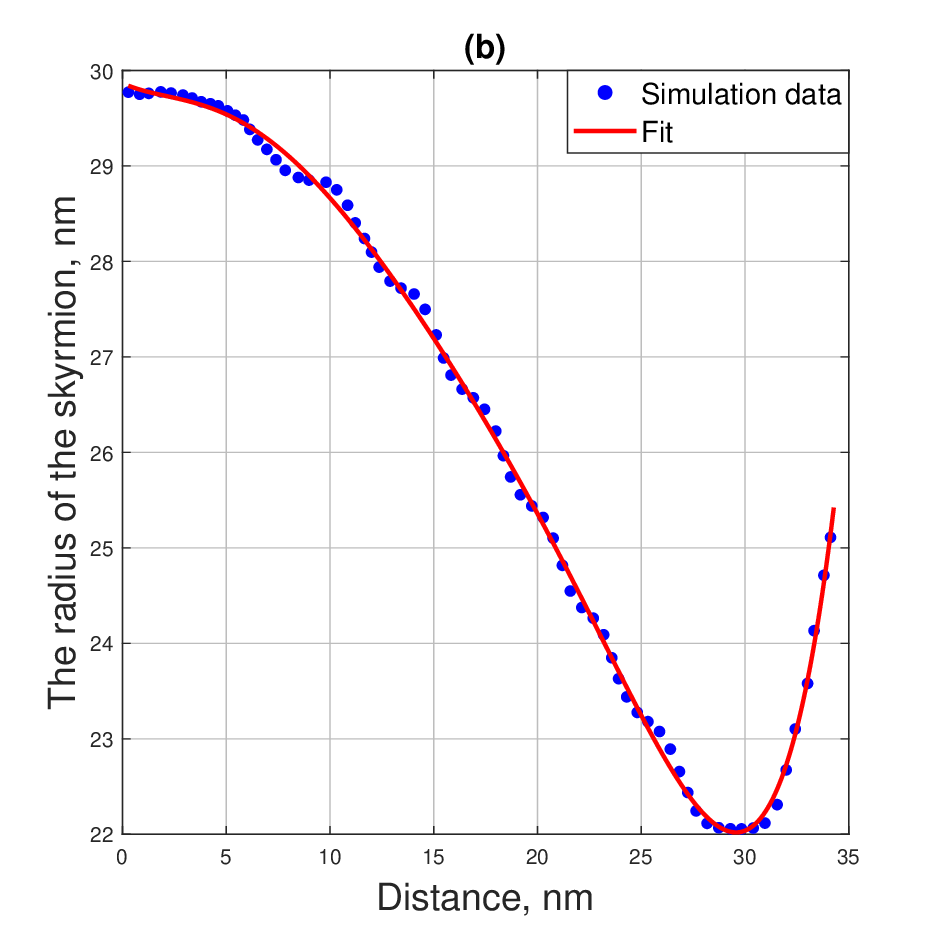}
  \caption{(a) Trajectory of the skyrmion center in the nanocylinder driven by
  a CIP current applied along the $x$ axis. (b) Skyrmion radius as a function
  of the displacement of its center $|\bm{R}|$.}
  \label{fig:trajectory}
\end{figure}

When the skyrmion is displaced (Fig.~\ref{fig:trajectory}a) from the equilibrium position located at the center of the cylinder under a direct spin-polarized electric current flowing along the $x$ axis, it becomes deformed, that is, its radius changes (Fig.~\ref{fig:trajectory}b), because to the interaction with the magnetization inhomogeneity at the sample boundary, as demonstrated by simulations in Boris Computational Spintronics software package~\cite{ref46} with the parameters listed in Appendix~\ref{app:params}.

We account for this deformation by substituting~\eqref{eq:weighted} into the magnetic energy functional $E$ and minimizing with respect to the parameters $\varepsilon$ and $\delta$, which characterize the skyrmion size. Thus, if the magnitude of the bias magnetic field $\bm{B}_{\mathrm{ext}}$ is fixed, then for a given skyrmion position $\bm{R}$ the parameters $\varepsilon$ and $\delta$ are found by minimizing the functional $E$, written as
\begin{equation}
  E\bigl[\varepsilon,\delta\bigr]
  = \int \bigl(\rho_{\mathrm{anis}} + \rho_{\mathrm{demag}}
    + \rho_{\mathrm{ex}} + \rho_{\mathrm{DMI}}
    + \rho_{\mathrm{Zee}}\bigr)\,dV,
  \label{eq:energy}
\end{equation}
which is defined as the integral over the sample volume relevant contributions: the anisotropy energy density $\rho_{\mathrm{anis}} = -K_u(\bm{M}\cdot\mathbf{e}_a)^2/M_s^2$, where $\mathbf{e}_a$ is the anisotropy unit vector, directed perpendicular to the plane of the nanodot; the demagnetizing energy density $\rho_{\mathrm{demag}} = -\mu_0\bm{M}\cdot\bm{H}_{\mathrm{demag}}/2$, where the demagnetizing field can be found as $\bm{H}_{\mathrm{demag}}(\bm{r}) = -\int \widehat{N}(\bm{r}' - \bm{r})\,\bm{M}(\bm{r}')\, d\bm{r}'$, where $\widehat{N}$ is the demagnetizing tensor; the exchange energy density $\rho_{\mathrm{ex}} = A_{\mathrm{ex}}\bigl[(\partial\bm{M}/\partial x)^2 + (\partial\bm{M}/\partial y)^2 + (\partial\bm{M}/\partial z)^2\bigr]/M_s^2$; the Dzyaloshinskii--Moriya energy density $\rho_{\mathrm{DMI}} = -\mu_0\bm{M}\cdot\bm{H}_{\mathrm{DMI}}/2$, where the field of the DMI is given by $\bm{H}_{\mathrm{DMI}} = 2D_{\mathrm{DMI}} \bigl[(\nabla\cdot\bm{M})\mathbf{e}_z - \nabla M_z\bigr]/\mu_0 M_s^2$; and the Zeeman energy density $\rho_{\mathrm{Zee}} = -\bm{M}\cdot\bm{B}_{\mathrm{ext}}$, where $\bm{B}_{\mathrm{ext}}$ is the bias magnetic field applied perpendicular to the film plane, $\bm{B}_{\mathrm{ext}} = (0,0,B_0)$.

\begin{figure}[t]
  \includegraphics[width=1\linewidth]{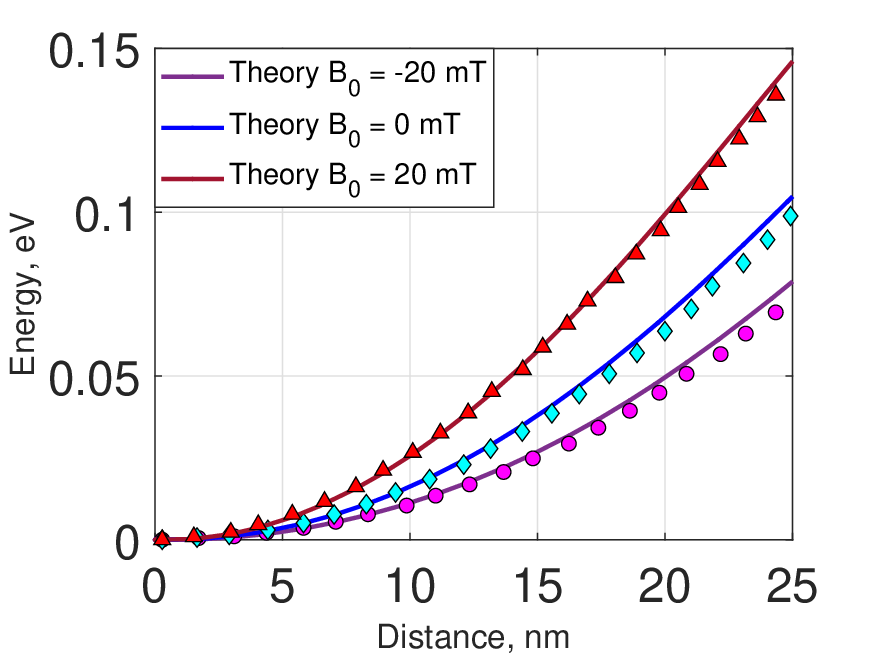}
  \caption{System energy $E$ as a function of the skyrmion displacement
  $|\bm{R}|$. Solid lines show the results of the theoretical calculations
  based on minimization of the magnetic energy functional. Circles represent
  the data obtained from micromagnetic simulations.}
  \label{fig:energy}
\end{figure}

A comparison of the resulting theoretical dependence of the energy $E(\bm{R})$ on the skyrmion displacement $|\bm{R}|$ with the results of micromagnetic simulations for various magnitudes of the bias magnetic field $\bm{B}_{\mathrm{ext}}$ is presented in Fig.~\ref{fig:energy}. For different magnitudes and signs of $B_0$ the relative difference between the theoretical results and the simulation data does not exceed 10\%. Note that, although a full energy minimization could be performed for every value of $|\bm{R}|$ in the subsequent calculations, a different approach is more convenient. For a given value of $B_0$ and a certain grid of values $\{\bm{R}_i\}$ we determine $\varepsilon(\{\bm{R}_i\})$ and $\delta(\{\bm{R}_i\})$ by minimization. The grid functions obtained in this way are then fitted by third-order polynomials in $|\bm{R}|^2$. The dependences $\varepsilon(|\bm{R}|)$ and $\delta(|\bm{R}|)$ can then be treated as continuous functions. Details of this fitting procedure are given in Appendix~\ref{app:minim}.

\section{SKYRMION DYNAMICS WITH DEFORMATIONS}
\label{sec:dynamics}

In the previous section we considered the calculation of the system energy for a skyrmion displaced from its equilibrium position, taking its deformation into account. We now proceed to the description of the dynamics. A popular method for the theoretical study of the motion of inhomogeneous magnetization states using the Thiele model~\cite{ref48}. Although this model is very convenient, it is not directly applicable to the case under study, because the Thiele equation is commonly derived within the rigid-object approximation. It is thus assumed that the non-trivial magnetization configuration under consideration undergoes no deformation. Nevertheless, Refs.~\cite{ref49,ref50,ref51} present a so-called generalized Thiele equation, which makes it possible to account for changes in the parameters of the inhomogeneity, including its characteristic size. We use this model, bearing in mind that the magnetization vector $\bm{M}$ depends on the spatial coordinates and on the skyrmion position, and that its arguments are independent, that is, $\bm{M}(\bm{r},\bm{R}(t))$.

We write the Landau--Lifshitz equation with Gilbert damping (LLG), supplemented by the Zhang--Li terms that arise when an electric current is passed through the nanocylinder, in the following form~\cite{ref52}
\begin{equation}
  \begin{gathered}
    \frac{d\bm{M}}{dt} =
      -\gamma\,\bm{M}\times\bm{H}^{\mathrm{eff}}
      + \frac{\alpha}{M_s}\bm{M}\times\frac{d\bm{M}}{dt}
      + \bm{T}_{\mathrm{STT}},\\[4pt]
    \bm{T}_{\mathrm{STT}} =
      \left(\bm{u}\frac{\partial}{\partial\bm{r}}\right)\bm{M}
      - \frac{\beta}{M_s}\bm{M}\times
        \left(\bm{u}\frac{\partial}{\partial\bm{r}}\right)\bm{M},
  \end{gathered}
  \label{eq:llg}
\end{equation}
where $\gamma$ is the gyromagnetic ratio, $\bm{H}^{\mathrm{eff}}$ is the
effective magnetic field, $\alpha$ is the Gilbert damping constant,
$\bm{u} = \mu_B p\,\bm{j}/eM_s(1+\beta^2)$ is the effective drift
velocity of the spin current, $\bm{j}$ is the current density, $\beta$ is
the non-adiabaticity parameter of the spin current, $\mu_B$ is the Bohr
magneton, $e$ is the elementary charge and $p$ is the polarization of the
spin-polarized current. The LLG equation describes the time evolution of the
magnetization vector and can be rewritten in the form
\begin{equation}
  \label{eq:llg2}
  \begin{split}
    &-\frac{1}{M_s^2}\bm{M}\times\frac{d\bm{M}}{dt}
    - \frac{\alpha}{M_s}\frac{d\bm{M}}{dt}
    + \gamma\bm{H}^{\mathrm{eff}} \\
    &+ \frac{1}{M_s^2}\bm{M}\times
      \left(\bm{u}\frac{\partial}{\partial\bm{r}}\right)\bm{M}
    + \frac{\beta}{M_s}
      \left(\bm{u}\frac{\partial}{\partial\bm{r}}\right)\bm{M} = 0.
  \end{split}
\end{equation}

Writing the $i$-th component of Eq.~\eqref{eq:llg2}, multiply it by
$\partial M_i/\partial R_j$ and sum the resulting expression over the index $i$ gives
\begin{subequations}
  \label{eq:proj}
  \begin{gather}
    f_j^{\,a,(1)} + f_j^{\,a,(2)} + f_j^{\,g,(1)} + f_j^{\,g,(2)} + f_j^{\,r}= 0,
    \label{eq:proj-b2}\\[4pt]
    f_j^{\,a,(1)} = -\frac{\alpha}{M_s}\frac{dM_i}{dt}
      \frac{dM_i}{dR_j},
    \label{eq:proj-c1}\\[4pt]
    f_j^{\,a,(2)} = \frac{\beta}{M_s}
      \left(\bm{u}\frac{\partial}{\partial\bm{r}}\right)M_i
      \frac{dM_i}{dR_j},
    \label{eq:proj-a}\\[4pt]
    f_j^{\,g,(1)} = -\frac{1}{M_s^2}e_{ilk}M_l
      \frac{d M_k}{dt}\frac{\partial M_i}{\partial R_j},
    \label{eq:proj-b1}\\[4pt]
    f_j^{\,g,(2)} = \frac{1}{M_s^2}e_{ilk}M_l
      \left(\bm{u}\frac{\partial}{\partial\bm{r}}\right)M_k
      \frac{\partial M_i}{\partial R_j},
    \label{eq:proj-c2}\\[4pt]
    f_j^{\,r} = \gamma H_i^{\mathrm{eff}}\frac{\partial M_i}{\partial R_j}
      = -\frac{\gamma}{\mu_0}\frac{\delta E}{\delta M_i}
        \frac{\partial M_i}{\partial R_j}.
    \label{eq:proj-d}
  \end{gather}
\end{subequations}
where, in taking the time derivative $t$, we considered that
$\bm{r} \neq \bm{r}(t)$ and $\bm{R}(t)$ are independent and $d{\bm{M}}/dt=\partial{\bm{M}}/\partial{\bm{R}}\cdot d{\bm{R}}/dt$; the energy
is defined by Eq.~\eqref{eq:energy}, and $e_{ilk}$ is the Levi-Civita symbol.
Let us consider separately the term $f_j^{\,a,(1)}$, where we replaced the indexes $i \leftrightarrow k$
\begin{subequations}
  \label{eq:dissip}
  \begin{align}
     f_j^{\,a,(1)} = -\frac{\alpha}{M_s}\frac{\partial M_k}{\partial R_j}
       \frac{\partial M_k}{\partial R_i}\frac{dR_i}{dt}
     = \alpha d_{ji}^{(1)}\frac{dR_i}{dt},
    \label{eq:dissip-a}
  \end{align}
where $d_{ji}^{(1)}$ is a component of the symmetric dissipation tensor. In a
similar way we obtain
  \begin{align}
     f_j^{\,a,(2)} = \frac{\beta}{M_s}\frac{\partial M_k}{\partial r_i}
       \frac{\partial M_k}{\partial R_j}u_i
     = \beta d_{ji}^{(2)}u_i.
    \label{eq:dissip-b}
  \end{align}
\end{subequations}

Now consider the term $f_j^{\,g}$, where the index $i$ is replaced by $m$ and the cyclic permutation of indices $(m,l,k) \to (l,m,k)$ is performed.
The factors multiplying $dR_i/dt$ form an
antisymmetric tensor $g_{ji}^{(1)}$, which can be represented by a vecto
\begin{subequations}
  \label{eq:gyro}
  \begin{gather}
    f_j^{\,g,(1)} = g_{ji}^{(1)}\frac{dR_i}{dt} = e_{jik}g_i^{(1)}\frac{dR_k}{dt},
      \label{eq:gyro-b} \\[4pt]
    g_i^{(1)} = -\frac{1}{2}e_{ijk}g_{ji}^{(1)}
      = -\frac{1}{2M_s^2}e_{ijk}e_{mnp}M_m
        \frac{\partial M_n}{\partial R_j}\frac{\partial M_p}{\partial R_k}.
      \label{eq:gyro-c}
  \end{gather}
The vector composed of the $g_i^{(1)}$ is called the gyrovector. We
proceed in a similar way
\label{eq:gyro2}
  \begin{gather}
    f_j^{\,g,(2)} = g_{ji}^{(2)}u_i = e_{jik}g_i^{(2)}u_k,
      \label{eq:gyro-e} \\[4pt]
    g_i^{(2)} = -\frac{1}{2}e_{ijk}g_{ji}^{(2)}
      = \frac{1}{2M_s^2}e_{ijk}e_{mnp}M_m
        \frac{\partial M_n}{\partial R_j}\frac{\partial M_p}{\partial r_k}.
      \label{eq:gyro-f}
  \end{gather}
\end{subequations}

From Eq.~\eqref{eq:proj-d}, the variational derivative of the energy $E$ with
respect to $\bm{M}$ can be represented as
\begin{equation}
  \label{eq:vardev}
  \begin{aligned}
    \frac{\delta E}{\delta \bm{M}}
    &= \frac{\partial\rho}{\partial\bm{M}}
    - \sum_i \frac{\partial}{\partial r_i}
      \left(\frac{\partial\rho}
        {\partial\bigl(\partial\bm{M}/\partial r_i\bigr)}\right) \\
    & - \sum_j \frac{\partial}{\partial R_j}
      \left(\frac{\partial\rho}
        {\partial\bigl(\partial\bm{M}/\partial R_j\bigr)}\right).
  \end{aligned}
\end{equation}

Note that $\partial\rho/\partial(\partial\bm{M}/\partial R_j) = 0$, since
$\rho(\bm{M},\partial\bm{M}/\partial r_i)$ is a function only of the
magnetization vector $\bm{M}$ and of its derivatives with respect to the
spatial coordinates $\partial\bm{M}/\partial r_i$. Let us transform the
derivative of the energy density $\rho$ with respect to the vector $\bm{R}$
and integrate over the sample volume. We obtain
\begin{equation}
  \label{eq:drho}
  \begin{split}
    \int \frac{\partial\rho}{\partial\bm{R}}\,dV
    &= \int \sum_i \frac{\partial\rho}{\partial M_i}
      \frac{\partial M_i}{\partial\bm{R}}\,dV \\
    & + \int \sum_{i,j}
      \frac{\partial\rho}{\partial\bigl(\partial M_i/\partial r_j\bigr)}
      \frac{\partial}{\partial\bm{R}}
      \left(\frac{\partial M_i}{\partial r_j}\right)dV.
  \end{split}
\end{equation}

We now consider separately the second term on the right-hand side of
Eq.~\eqref{eq:drho}, having first interchanged the order of the partial
derivatives
\begin{multline}
    \frac{\partial\rho}{\partial\bigl(\partial M_i/\partial r_j\bigr)}
    \frac{\partial}{\partial r_j}
    \left(\frac{\partial M_i}{\partial\bm{R}}\right)
  =   \frac{\partial}{\partial r_j}
    \left(\frac{\partial\rho}
      {\partial\bigl(\partial M_i/\partial r_j\bigr)}
      \frac{\partial M_i}{\partial\bm{R}}\right)\\
  -  \frac{\partial}{\partial r_j}
    \left(\frac{\partial\rho}
      {\partial\bigl(\partial M_i/\partial r_j\bigr)}\right)
    \frac{\partial M_i}{\partial\bm{R}}.
  \label{eq:parts}
\end{multline}

The first term on the right-hand side of Eq.~\eqref{eq:parts} is reduced by the
Gauss--Ostrogradsky theorem
\begin{equation}
  \label{eq:gauss}
  \begin{gathered}
    \int \sum_{i,j} \frac{\partial}{\partial r_j}
      \left(\frac{\partial\rho}
        {\partial\bigl(\partial M_i/\partial r_j\bigr)}
        \frac{\partial M_i}{\partial\bm{R}}\right)dV \\
    = \int \sum_{i,j}
      \left(\frac{\partial\rho}
        {\partial\bigl(\partial M_i/\partial r_j\bigr)}
        \frac{\partial M_i}{\partial\bm{R}}\right)n_j\,dS.
  \end{gathered}
\end{equation}
where $\bm{n}$ is the outward normal to the surface $S$ of the sample.
Using the boundary conditions that arise from the DMI Eq.~\eqref{eq:bc}, the last integral vanishes. Then, for the integral of the
derivative of the energy density with respect to $\partial\rho/\partial\bm{R}$ Eq.~\eqref{eq:drho} we obtain the expression
\begin{subequations}
\begin{equation}
  \label{eq:dE}
  \begin{split}
    \int \frac{\partial\rho}{\partial\bm{R}}\,dV
    &= \int \sum_i \frac{\partial\rho}{\partial M_i}
       \frac{\partial M_i}{\partial\bm{R}}\,dV \\
    &- \sum_{i,j} \int \frac{\partial}{\partial r_j}
       \left(\frac{\partial\rho}
         {\partial\bigl(\partial M_i/\partial r_j\bigr)}\right)
       \frac{\partial M_i}{\partial\bm{R}}\,dV.
  \end{split}
\end{equation}
which coincides with Eq.~\eqref{eq:vardev} and can be rewritten as
  \begin{align}
    \frac{\partial E}{\partial\bm{R}}
    &= \int \frac{\partial\rho}{\partial\bm{R}}\,dV
     = \int \sum_i \frac{\delta E}{\delta M_i}
       \frac{\partial M_i}{\partial\bm{R}}\,dV.
    \label{eq:dE-b}
  \end{align}
\end{subequations}

Integrating Eq.~\eqref{eq:proj-a} over the sample volume $V$, we obtain
\begin{subequations}
  \label{eq:forces}
  \begin{align}
    F_j^{\,a} &= \int f_j^{\,a}\,dV \notag \\
    &= \alpha\int d_{ji}^{(1)}\,dV \frac{\partial R_i}{\partial t}
      + \beta\int d_{ji}^{(2)}\,dV\, u_i \notag \\
    &= \alpha D_{ji}^{(1)}\frac{dR_i}{dt} + \beta D_{ji}^{(2)}u_i,
    \label{eq:forces-a} \\[4pt]
    F_j^{\,g} &= \int f_j^{\,g}\,dV \notag \\
    &= e_{jik}\int g_i^{(1)}\,dV \frac{dR_k}{dt}
      + e_{jik}\int g_i^{(2)}\,dV\, u_k \notag \\
    &= e_{jik}G_i^{(1)}\frac{dR_k}{dt} + e_{jik}G_i^{(2)}u_k,
    \label{eq:forces-b} \\[4pt]
    F_j^{\,r} &= -\frac{\gamma}{\mu_0}
      \int \frac{\delta E}{\delta M_i}\frac{\partial M_i}{\partial R_j}\,dV
    = -\frac{\gamma}{\mu_0}\frac{\partial E}{\partial R_j},
    \label{eq:forces-c}
  \end{align}
\end{subequations}
and then we write the generalized Thiele equation in vector form, moving the
energy derivative to the right-hand side
\begin{equation}
  \alpha D^{(1)}\frac{d\bm{R}}{dt} + \beta D^{(2)}\bm{u}
  + \bm{G}^{(1)}\times\frac{d\bm{R}}{dt}
  + \bm{G}^{(2)}\times\bm{u}
  = \frac{\gamma}{\mu_0}\frac{\partial E}{\partial\bm{R}},
  \label{eq:thiele}
\end{equation}
in which, generally speaking, $D^{(1)}$, $D^{(2)}$, $\bm{G}^{(1)}$ and
$\bm{G}^{(2)}$ are functions of the position of the center of the
inhomogeneity $\bm{R}$. We assume that $\partial/\partial z = 0$ and
$\partial/\partial R_z = 0$ and therefore
$G_x^{(1,2)} = G_y^{(1,2)} = 0$. This approximation is valid for thin magnetic
films.

\section{HAMILTONIAN FORMALISM FOR THE GENERALIZED THIELE EQUATION}
\label{sec:hamiltonian}

Equation~\eqref{eq:thiele} is rather difficult to analyze directly. Therefore,
first, we simplify it by expanding the functions $D^{(1)}$, $D^{(2)}$,
$\bm{G}^{(1)}$ and $\bm{G}^{(2)}$ in a Taylor series in the variables
$R_x$ and $R_y$ near the point $(0,0)$ up to fourth order. We write the
general form of such expansion as
\begin{equation}
  \label{eq:taylor}
  \begin{split}
    \upsilon(R_x,R_y) &= \upsilon_0 + \upsilon_{10}R_x + \upsilon_{01}R_y \\
    &\quad + \upsilon_{20}R_x^2 + \upsilon_{11}R_xR_y + \upsilon_{02}R_y^2 + \ldots,
  \end{split}
\end{equation}
where $\upsilon_0 = \upsilon(0,0)$ is the value of arbitrary function in the absence
of a skyrmion displacement from the equilibrium position and
$\upsilon_{ij} = \partial^{i+j}\upsilon/
(\partial R_x^i \partial R_y^j)\big|_{(0,0)}$. The term on the right-hand side
of Eq.~\eqref{eq:thiele} is called the conservative force, since it is computed
as the gradient with respect to the coordinate of the center of the
inhomogeneity. We consider corresponding expansion for magnetic energy
\begin{equation}
  E \approx E_0 + \kappa|\bm{R}|^2 + \chi|\bm{R}|^4+ \ldots.
  \label{eq:Eexp}
\end{equation}
Here the stiffness coefficients $\kappa$ and $\chi$ depends on bias magnetic field.

Second, to describe the skyrmion dynamics we introduce the circular coordinates
$a = R_x - iR_y$ and $a^{*} = R_x + iR_y$. Note that, with this change of
variables, the Thiele equation can be reduced to the quasi-Hamiltonian form
so one can obtain
\begin{equation}
  \frac{da}{dt} = -i\frac{\partial H_a}{\partial a^{*}} + F_a + \mathcal{F}_a,
  \label{eq:ham}
\end{equation}
where $H_a(a,a^{*})$ is the Hamiltonian of the system; $F_a(a,a^{*})$ is the perturbing damping force proportional
to $\alpha$; and $\mathcal{F}_a(a,a^{*})$ is the perturbing force arising from
the Zhang--Li torque. The Hamiltonian function of the system then takes the form
\begin{multline}
  H_a = \omega|a|^2
  + \tfrac{1}{4}i\bigl(2A' - B'^{*}\bigr)a|a|^2
  + \tfrac{1}{4}i\bigl(B' - 2A'^{*}\bigr)a^{*}|a|^2\\
  + \tfrac{1}{6}i\bigl(3C' - K'^{*}\bigr)a^2|a|^2
  + \tfrac{1}{6}i\bigl(K' - 3C'^{*}\bigr)(a^{*})^2|a|^2
  + \tfrac{1}{2}F'|a|^4,
  \label{eq:Ha}
\end{multline}
whose coefficients are given in Appendix~\ref{app:coefficients}.

Using a canonical nonlinear transformation, one can eliminate the terms
$\sim a^3$ and $\sim a^4$ except for $|a|^4$. To do this, one has to use the
substitution
$a = c + \xi_1 c^2 + \xi_2|c|^2 + \xi_3 (c^{*})^2 + \eta_1 c^3
+ \eta_2 c|c|^2 + \eta_3 c^{*}|c|^2 + \eta_4 (c^{*})^3 + O(c^4)$,
which satisfies the canonicity condition
$\{a(c,c^{*}),a^{*}(c,c^{*})\} = 1$. Solving the system of algebraic
equations, one obtains the coefficients $\xi_i$ and $\eta_i$. As a result of
this substitution, the Hamiltonian of the system is rewritten as
\begin{equation}
  H_c = \omega|c|^2 + \frac{N}{2}|c|^4,
  \label{eq:Hc}
\end{equation}
where $N$ is the nonlinearity coefficient, determined by
\begin{equation}
  \label{eq:N}
  \begin{split}
    N &= F' + 2\omega\bigl(|\xi_1|^2 + |\xi_2|^2
      + 2\operatorname{Re}(\eta_2)\bigr)
    - 4\operatorname{Im}(A'\xi_2) \\
    &- \operatorname{Im}(B'\xi_1)
    - 2\operatorname{Im}(A'\xi_1^{*})
    - 2\operatorname{Im}(B'\xi_2^{*}).
  \end{split}
\end{equation}

The derivation of the perturbing force expressions is presented below.
Expanding in a series those terms on the right-hand side of
Eq.~\eqref{eq:ham} that arise from the Gilbert damping, and applying the
quasi-canonical nonlinear transformation, allows the dissipative perturbing
force for $c$ to be written as
\begin{equation}
  F_c = \Gamma_0 c + Q c|c|^2.
  \label{eq:Fc}
\end{equation}

For the perturbing force arising from the current passed through the sample,
after transformations analogous to those used to derive $F_c$, we retain only
one term. Note that keeping the terms proportional to the product of $j(t)$
with powers of $c$ and $c^{*}$ would make it possible to describe parametric
effects. Such effects are not considered in the present work. One
can write the dynamical equation for $c$ in the form
\begin{equation}
  \frac{dc}{dt} = -i\bigl(\omega + N|c|^2\bigr)c
  + \bigl(\Gamma_0 + Q|c|^2\bigr)c + g_{\varphi}\,j(t),
  \label{eq:dyn}
\end{equation}
where the coefficient $g_{\varphi}$ does not depend on the skyrmion position or
on the magnitude of the current density, but does depend on the direction in
which the current flows, through the angle $\varphi$ formed between the current
density and the $x$ axis.

The amplitude-frequency response of the oscillator obtained from micromagnetic simulations can be compared with the corresponding theoretical curves by considering Eq.~\eqref{eq:dyn} and seeking it is stationary points. This is described in more detail in Appendix~\ref{app:stat}.

\section{RESULTS}
\label{sec:results}

\begin{figure}[H]
  \includegraphics[width=1\linewidth]{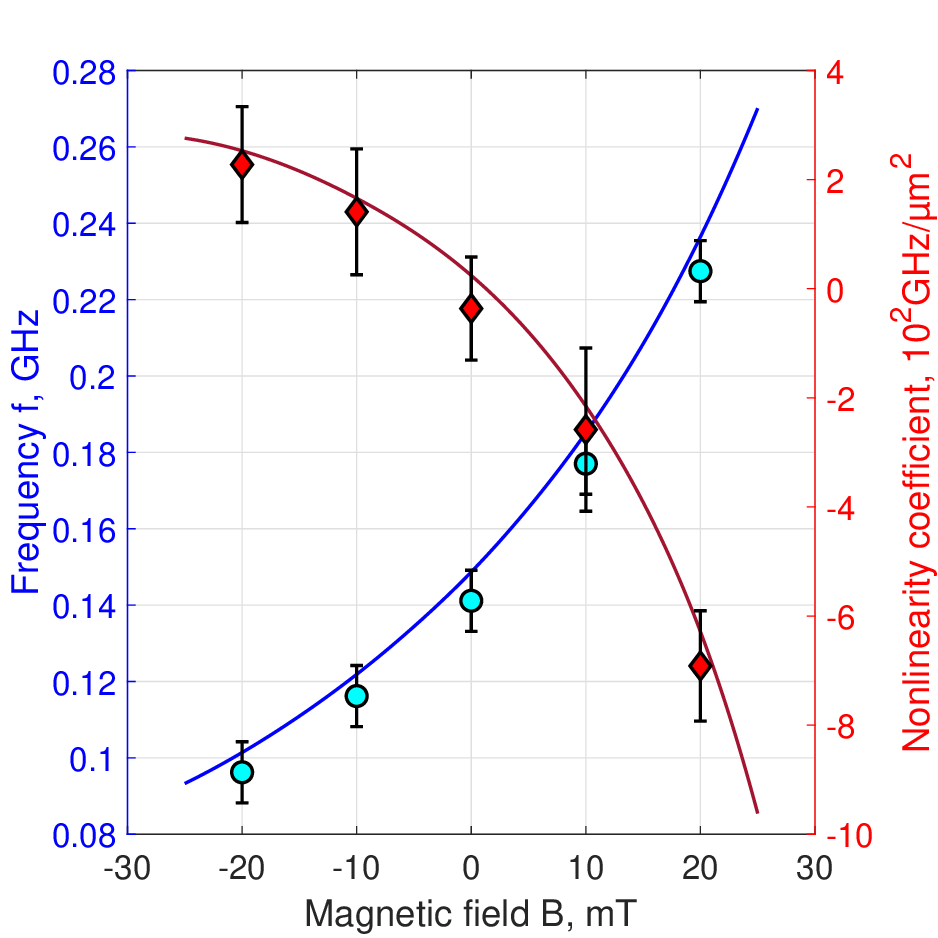}
  \caption{Frequency $f = \omega/2\pi$ (blue solid line) and nonlinearity
  coefficient $N$ (red dashed line) as functions of the bias
  magnetic field $B_0$ perpendicular to the plane of the nanocylinder.}
  \label{fig:freq}
\end{figure}

The dependence of the frequency $\omega$ and of the nonlinearity coefficient $N$ on the bias magnetic field $B_0$ can be explained as follows: the functions $\varepsilon$ and $\delta$ depend on the bias magnetic field and, consequently, so does the function $\bm{G}^{(1)}$, which contains there parameters. The largest contribution to the field dependence of these quantities comes from the energy-expansion coefficients $\kappa$ and $\chi$. Figure~\ref{fig:freq} shows the frequency
$f = \omega/2\pi$ and the nonlinearity coefficient $N$ (red dashed line) as
functions of the bias magnetic field $B_0$ perpendicular to the
plane of the nanocylinder, obtained theoretically and from simulations. As the field is varied, the nonlinearity coefficient $N$ is seen to reverse sign.

\begin{figure}
  \centering
  \begin{minipage}[c]{0.05\linewidth}
    \centering (a)
  \end{minipage}
  \begin{minipage}[c]{0.85\linewidth}
    \centering
    \includegraphics[width=\linewidth]{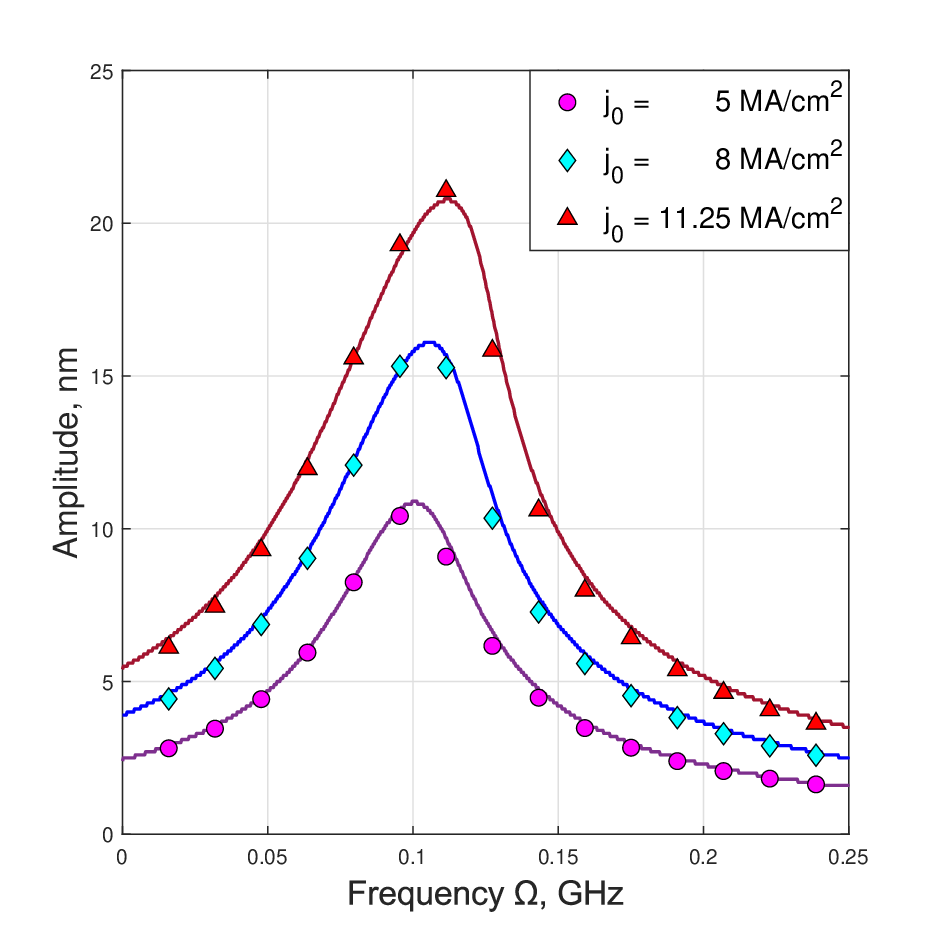}
  \end{minipage}\\[-6pt]
  \begin{minipage}[c]{0.05\linewidth}
    \centering (b)
  \end{minipage}
  \begin{minipage}[c]{0.85\linewidth}
    \centering
    \includegraphics[width=\linewidth]{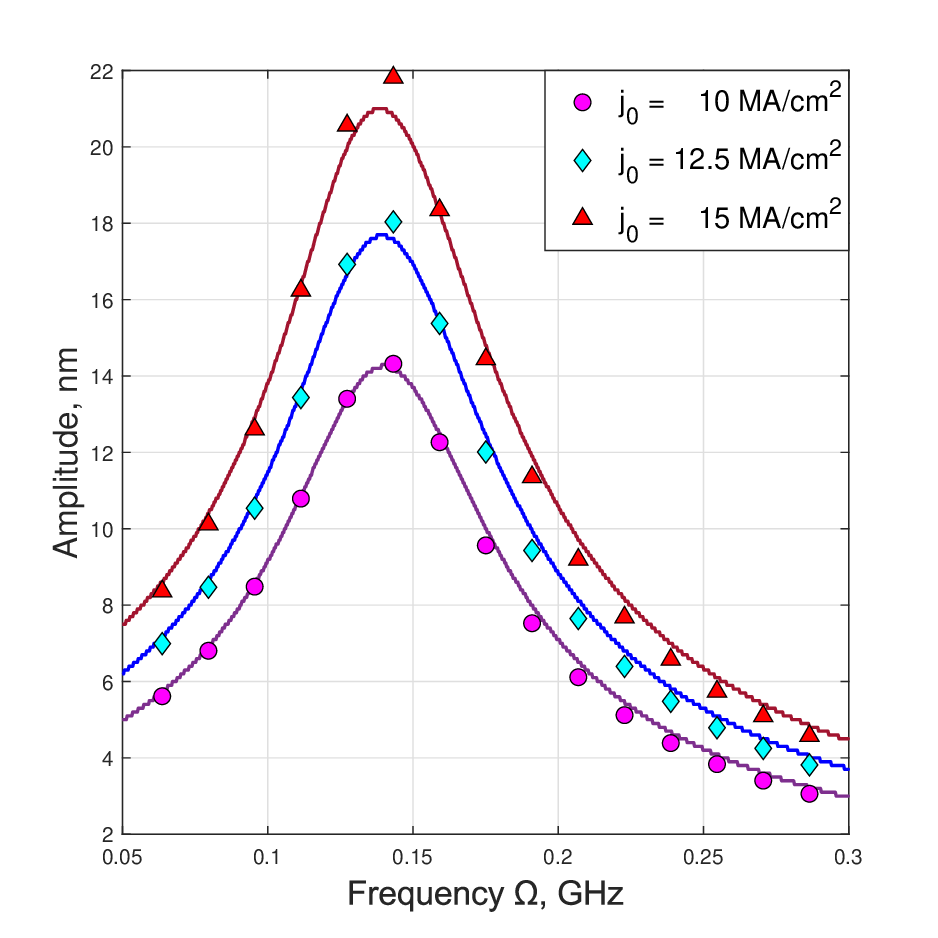}
  \end{minipage}\\[-6pt]
  \begin{minipage}[c]{0.05\linewidth}
    \centering (c)
  \end{minipage}
  \begin{minipage}[c]{0.85\linewidth}
    \centering
    \includegraphics[width=\linewidth]{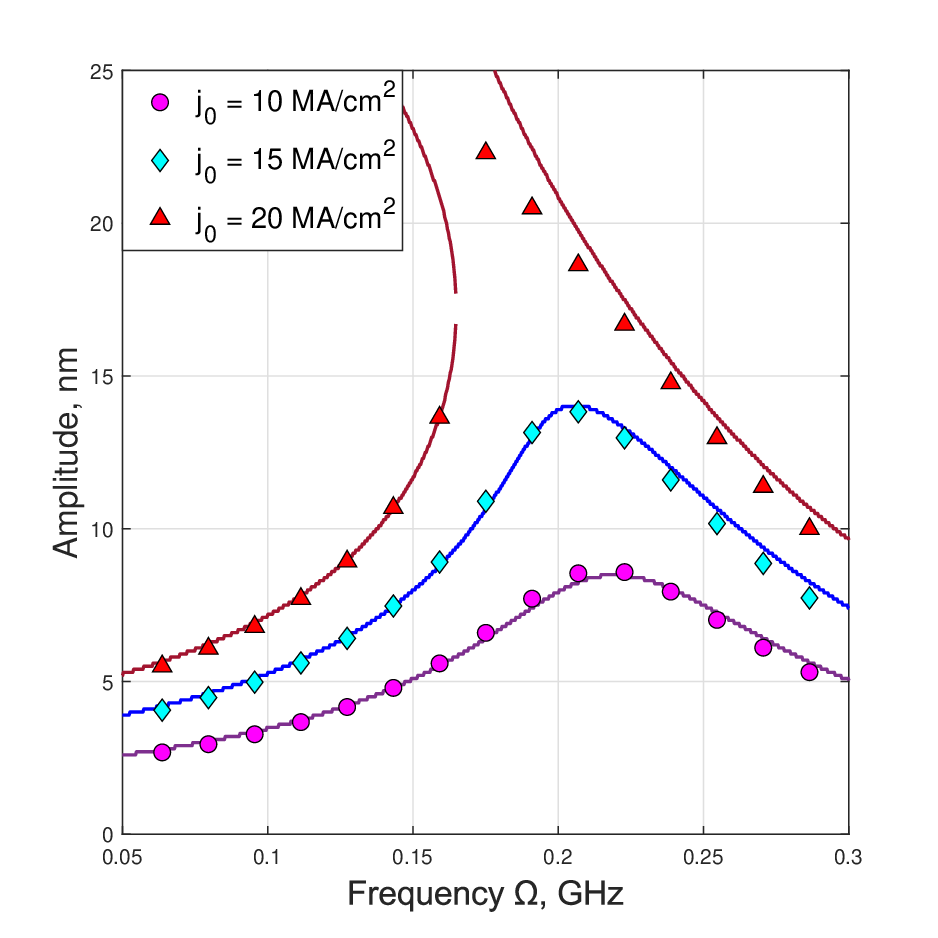}
  \end{minipage}
  \caption{Amplitude--frequency response of the skyrmion oscillator in a
  magnetic field $B_0$ of magnitude (a) $-20$~mT, (b) $0$~T, (c) $20$~mT.}
  \label{fig:afr}
\end{figure}

The sign reversal of $N$ leads to a qualitative change in the amplitude-frequency response of the oscillator. The dependences of the steady-state oscillation amplitude on the frequency of the current-density modulation for various values of the bias field $B_0$ are presented in Fig.~\ref{fig:afr}, which shows both the simulated data points and the theoretical fit curves. As the field $B_0$ is varied, both the resonance frequency $\omega_{\mathrm{res}}$ and the shape of the resonance peak change accordingly.

\begin{figure}[t]
  \includegraphics[width=1\linewidth]{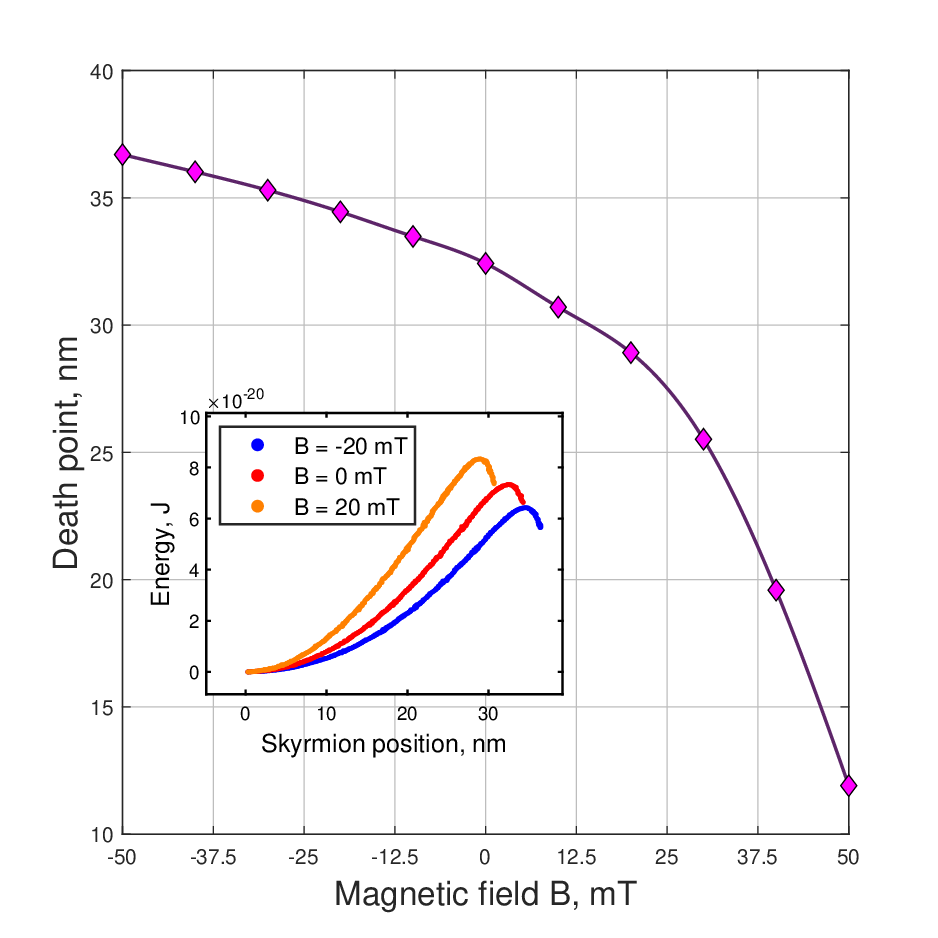}
  \caption{Dependence of the death point (the point at which the derivative of
  the energy with respect to the coordinate of the skyrmion center changes
  sign) on the magnitude of the applied magnetic field. Inset: the
  energy as a function of the skyrmion displacement from the equilibrium
  position at different fields (the blue curve corresponds to a magnetic field
  $B_0 = -20$~mT, the red one to $B_0 = 0$~mT and the orange one to
  $B_0 = 20$~mT).}
  \label{fig:death}
\end{figure}

The dependence of the frequency $\omega$ and of the nonlinear frequency-shift
coefficient $N$ on the magnitude and sign of $B_0$ can be explained as follows.
The inset of Fig.~\ref{fig:death} shows the energy as a function of the
skyrmion displacement from the center of the nanocylinder for various bias
fields. One can see that there exists a value $R = R_{\mathrm{dp}}$ at which
$\partial E/\partial R = 0$. Moreover, for $R > R_{\mathrm{dp}}$ this derivative is
negative, whereas for $R < R_{\mathrm{dp}}$ the sign of $\partial E/\partial R$ is
positive. Consequently, once the energy barrier $E(R_{\mathrm{dp}})$ is overcome,
expulsion from the sample becomes energetically favourable for the skyrmion, as
has been observed for magnetic nanostripes in Ref.~\cite{ref29}. Thus, there is a certain critical displacement $R = R_{\mathrm{dp}}$ of the skyrmion from the center of the nanocylinder, overcoming which leads to the destruction of the skyrmion pattern.

As the magnitude of the bias magnetic field is changed, a larger number of
magnetic moments align with the field direction, since such an orientation is
energetically more favourable. This alters the skyrmion size and the energy,
including the coefficients in its expansion~\eqref{eq:Eexp}. The value of
$R_{\mathrm{dp}}$ shifts with the field (Fig.~\ref{fig:death}), so that the magnitudes
of the coefficients $\kappa$ and $\chi$ change, and the sign of $\chi$ changes
as well. As noted above, it is precisely $\chi$ that strongly affects the
magnitude of the nonlinearity coefficient. A magnetic field directed opposite
to the magnetization at the skyrmion core reduces the skyrmion size and also
weakens its interaction with the magnetization distribution at the boundary,
which enlarges the region available for skyrmion motion without annihilation at
the boundary; a field directed along the magnetization at the skyrmion core, by
contrast, increases its size and reduces the allowed region of motion.

\section{CONCLUSION}
\label{sec:conclusion}

We have formulated the derivation of the generalized Thiele model. In the present work the magnetization of the sample depended on six
independent coordinates in the general case, but the derivation remains valid
for a larger number of variables. The key point is the dependence of the energy
and of the energy density on the components of the magnetization vector and on
their derivatives. With this in mind, using the basic formulas of the calculus
of variations, we have explicitly demonstrated how the term equal to the
gradient of the energy with respect to the position of the skyrmion center is
obtained.

Together with the domain wall, we have taken into account the deformation of
the skyrmion as it is displaced from its equilibrium position towards the
sample boundary under the action of the driving force. This deformation arises
from the interaction of the skyrmion with the inhomogeneous magnetization
distribution at the boundary, which is induced by the Dzyaloshinskii--Moriya
interaction~\cite{ref29}. This interaction can be viewed as a resistance whole magnetization pattern to motion motion of skyrmion guiding center towards the boundary, as a result of which its size, and
hence its energy, decreases. At a fixed field, the change in size is accounted
for through the functional dependence of the minimization parameters on the
position of the skyrmion center, which we determined by minimizing the energy
with respect to these parameters.

We have demonstrated the possibility of controlling the frequency
and the nonlinearity coefficient by means of bias magnetic field (Fig.~\ref{fig:afr}). This is more feasible
technically than controlling the nonlinearity by means of the angle
between the field and plane, as proposed for STNO uniform mode
in Ref.~\cite{ref53,slavin2008excitation}. The results of this study broaden the prospects for using skyrmion-based STNOs as tunable spintronic elements for applications such as neuromorphic computing.

\begin{acknowledgments}
This work was carried out with support of the Russian Science Foundation (Project No. 25-79-20053).
\end{acknowledgments}

\appendix

\section{GEOMETRY AND MATERIAL PARAMETERS}
\label{app:params}

The micromagnetic simulations were performed with the Boris software
package~\cite{ref46}. The radius of the thin nanocylinder is $R_0 = 50$~nm and
its thickness is $l_z = 0.8$~nm, so the system under consideration can be
regarded as a single layer of magnetic material. The computational mesh in
Boris is $250 \times 250 \times 1$. The simulations used magnetic parameters
typical of the multilayers $\mathrm{Pt/Co/Ir}$: the
exchange stiffness constant $A_{\mathrm{ex}} = 16\times10^{-12}$~J/m, the
Dzyaloshinskii--Moriya constant $D_{\mathrm{DMI}} = 1.5$~mJ/m$^2$,
$K_u = 0.717\times10^{6}$~J/m$^3$ is the perpendicular anisotropy constant, the
saturation magnetization $M_s = 0.956\times10^{6}$~A/m and the Gilbert damping
constant $\alpha = 0.2$.

\section{MINIMIZATION OF THE ENERGY FUNCTIONAL}
\label{app:minim}

Minimizing the total micromagnetic energy for each individual value of the
skyrmion displacement $|\bm{R}|$ requires substantial computational
resources, especially when the calculation has to be repeated many times for
different external fields. Instead, we employ a parametric approach in which
the skyrmion shape is described by two key geometrical parameters: the size
coefficient $\varepsilon$ and the steepness parameter $\delta$, which govern
the radial magnetization profile. For a fixed magnitude of the applied magnetic
field $B_0$ a two-dimensional grid of values
$\{\varepsilon_i,\delta_j\}$ is constructed over ranges that are selected
separately for each field so as to guarantee that the energy minimum falls
inside the grid and that cover all possible skyrmion states within the range of
displacements studied.

For each discrete displacement from the set $\{\bm{R}_i\}$  we construct trial magnetization profiles
$\bm{m}(r,\varepsilon_i,\delta_j,\bm{R})$, using the analytical model
described in the main text. The total energy
$E(\varepsilon_i,\delta_j,\bm{R})$ for any pair of parameters can be
computed by numerically integrating the energy density over the sample volume.
Minimizing this function with respect to $\varepsilon_i$ and $\delta_j$, we
find the optimal values $\varepsilon_{\mathrm{opt}}$ and
$\delta_{\mathrm{opt}}$ corresponding to the lowest energy for the given
displacement. Thus, for each position of the skyrmion center we obtain a triple
of quantities: the minimum energy $E_{\min}(|\bm{R}|)$ together with
$\varepsilon(|\bm{R}|) = \varepsilon_{\mathrm{opt}}$ and
$\delta(|\bm{R}|) = \delta_{\mathrm{opt}}$.

The resulting discrete dependences $\varepsilon(|\bm{R}|)$ and
$\delta(|\bm{R}|)$ are fitted by third-degree polynomials in
$|\bm{R}|^2$ using the least-squares method. This allows these
parameters to be represented as continuous functions of the displacement
\begin{equation}
  \begin{gathered}
    \varepsilon(|\bm{R}|) \approx \varepsilon_0
      + \varepsilon_1|\bm{R}|^2 + \varepsilon_2|\bm{R}|^4
      + \varepsilon_3|\bm{R}|^6,\\[4pt]
    \delta(|\bm{R}|) \approx \delta_0
      + \delta_1|\bm{R}|^2 + \delta_2|\bm{R}|^4
      + \delta_3|\bm{R}|^6,
  \end{gathered}
  \label{eq:approx}
\end{equation}
where the coefficients are determined separately for each value of the magnetic
field $B_0$. Thanks to this approximation, we can compute the
energy of the system for an arbitrary displacement $|\bm{R}|$ as
$E(|\bm{R}|) = E\bigl(\varepsilon(|\bm{R}|),
\delta(|\bm{R}|),|\bm{R}|\bigr)$, which considerably simplifies the calculations.

\section{COEFFICIENTS OF THE HAMILTONIAN}
\label{app:coefficients}

The coefficients used in the expression for the Hamiltonian~\eqref{eq:Ha}
are given by
\begin{equation}
  \omega = -\frac{2\widetilde{\kappa}}{G_{z_0}^{(1)}},
  \label{eq:omega_app}
\end{equation}
\begin{equation}
  A' = \frac{\widetilde{\kappa}\bigl(G_{z_{01}}^{(1)} - iG_{z_{10}}^{(1)}\bigr)}
  {\bigl(G_{z_0}^{(1)}\bigr)^2},
  \label{eq:Aprime_app}
\end{equation}
\begin{equation}
  B' = -\frac{\widetilde{\kappa}\bigl(G_{z_{01}}^{(1)} + iG_{z_{10}}^{(1)}\bigr)}
  {\bigl(G_{z_0}^{(1)}\bigr)^2},
  \label{eq:Bprime_app}
\end{equation}
\begin{equation}
\begin{split}
  C' {} & = \frac{\widetilde{\kappa}}{2\bigl(G_{z_0}^{(1)}\bigr)^3}
  \Bigl[G_{z_0}^{(1)}G_{z_{11}}^{(1)} - 2G_{z_{10}}^{(1)}G_{z_{01}}^{(1)} \\
  & - i\Bigl(G_{z_0}^{(1)}\bigl(G_{z_{20}}^{(1)} - G_{z_{02}}^{(1)}\bigr)
  - \bigl(G_{z_{10}}^{(1)}\bigr)^2 + \bigl(G_{z_{01}}^{(1)}\bigr)^2\Bigr)\Bigr]
\end{split}
\label{eq:Cprime_app}
\end{equation}
\begin{equation}
\begin{split}
  F' {} & = -\frac{1}{\bigl(G_{z_0}^{(1)}\bigr)^3}
  \Bigl[4\bigl(G_{z_0}^{(1)}\bigr)^2\widetilde{\chi}
  + \widetilde{\kappa}\Bigl(-G_{z_0}^{(1)}G_{z_{20}}^{(1)} \\
  & - G_{z_0}^{(1)}G_{z_{02}}^{(1)}
  + \bigl(G_{z_{10}}^{(1)}\bigr)^2 + \bigl(G_{z_{01}}^{(1)}\bigr)^2\Bigr)\Bigr]
\end{split}
\label{eq:Fprime_app}
\end{equation}
\begin{equation}
\begin{split}
  K' {} & = -\frac{\widetilde{\kappa}}{2\bigl(G_{z_0}^{(1)}\bigr)^3}
  \Bigl[G_{z_0}^{(1)}G_{z_{11}}^{(1)} - 2G_{z_{10}}^{(1)}G_{z_{11}}^{(1)} \\
  & + i\Bigl(G_{z_0}^{(1)}\bigl(G_{z_{20}}^{(1)} - G_{z_{02}}^{(1)}\bigr)
  - \bigl(G_{z_{10}}^{(1)}\bigr)^2 + \bigl(G_{z_{01}}^{(1)}\bigr)^2\Bigr)\Bigr]
\end{split}
\label{eq:Kprime_app}
\end{equation}
where $\widetilde{\kappa} = \gamma\kappa/\mu_0$ and $\widetilde{\chi} = \gamma\chi/\mu_0$.

\section{STATIONARY POINTS}
\label{app:stat}

Let us derive the equation that determines the steady-state oscillation
amplitude of the system. To this end, in Eq.~\eqref{eq:dyn} we use the
substitution
\begin{equation}
  c(t) = u(t)\,e^{-i\left(\Omega t + \psi(t)\right)},
  \label{eq:sub}
\end{equation}
where $\Omega$ is the frequency of the alternating current
$j(t) = j_0\cos(\Omega t)$, $u(t)$ is the real slowly varying amplitude and
$\psi(t)$ is the phase. Note also that the coefficient $g_{\varphi}$ is complex
in the general case, so we rewrite it as
$g_{\varphi} = |g_{\varphi}|e^{i\theta}$. We separate the expression obtained
after the substitution into its real and imaginary parts, discarding the overtones. So we obtain
\begin{subequations}
  \label{eq:slow}
  \begin{align}
    \frac{du}{dt} &= \Gamma_0 u + Q u^3
      + \frac{|g_{\varphi}|}{2}j_0\cos(\psi + \theta),
    \label{eq:slow-a}\\[4pt]
    u\frac{d\psi}{dt} + \Omega u &= \omega u + N u^3
      - \frac{|g_{\varphi}|}{2}j_0\sin(\psi + \theta).
    \label{eq:slow-b}
  \end{align}
\end{subequations}

To find the steady-state regime, we set the derivatives $du/dt$ and $d\psi/dt$ equal to zero. Under this condition
Eqs.~\eqref{eq:slow-a} and \eqref{eq:slow-b} express $cos(\psi+\theta)$ and $sin(\psi+\theta)$ respectively, in terms of $u$. Squaring these two expressions and adding them together eliminates the phase $\psi+\theta$ and yields the equation for the amplitude
\begin{equation}
  \bigl(\Gamma_0 u + Q u^3\bigr)^2
  + \bigl((\omega - \Omega)u + N u^3\bigr)^2
  = \left(\frac{|g_{\varphi}|}{2}j_0\right)^2.
  \label{eq:stat}
\end{equation}

The resonance frequency $\Omega_{\mathrm{res}}$ corresponds to the maximum of $p=u^2$ with respect to $\Omega$ for a fixed current amplitude $j_0$, i.e., to $d(p)/d\Omega=0$. Differentiating Eq.~\eqref{eq:stat} with respect to $\Omega$ and applying this condition yields

\begin{equation}
  \Omega_{\mathrm{res}} = \omega + Np.
  \label{eq:resfreq}
\end{equation}

Equation~\eqref{eq:stat} is an algebraic equation of third order in $p$,
which in the general case may have up to three positive roots, corresponding to
bistable behaviour of the nonlinear oscillator. Solution of
Eq.~\eqref{eq:stat} makes it possible to find the positive roots $p$ for various values of the frequency \(\Omega\) of the in-plane
spin-polarized current and its density \(j_{0}\).

	\bibliography{refs}
	
\end{document}